%% file: arxiv.tex
\documentclass[10pt, a4paper, twocolumn]{article} % 10pt font size (11 and 12 also possible), A4 paper (letterpaper for US letter) and two column layout (remove for one column)

\input{assets/structure}

\fancypagestyle{firstpage}{ % Page style for the first page with the title
	\fancyhf{}
	\fancyfoot[C]{\small{This article has been accepted for publication in the IEEE Network Magazine and is subject to IEEE Copyright.}}
}

\title{Towards Blockchain-based Trust and Reputation Management for Trustworthy 6G Networks}

\author{
	\authorstyle{Guntur Dharma Putra\textsuperscript{1,2}, Volkan Dedeoglu\textsuperscript{3}, Salil S Kanhere\textsuperscript{1,2} and Raja Jurdak\textsuperscript{4}}
	\newline\newline
	\textsuperscript{1}\institution{School of Computer Science and Engineering, UNSW Sydney, Australia}\\
	\textsuperscript{2}\institution{Cyber Security Cooperative Research Centre (CSCRC), Australia}\\
	\textsuperscript{3}\institution{Commonwealth Scientific and Industrial Research Organisation's Data61, Pullenvale, Australia}\\
	\textsuperscript{4}\institution{School of Computer Science, Queensland University of Technology (QUT), Brisbane, Australia}
}

\date{\vspace{-7ex}}

\begin{document}

\maketitle

\thispagestyle{firstpage}

\lettrineabstract{6G is envisioned to enable futuristic technologies, which exhibit more complexities than the previous generations, as it aims to bring connectivity to a large number of devices, many of which may not be trustworthy. Proper authentication can protect the network from unauthorized adversaries. However, it cannot guarantee in situ reliability and trustworthiness of authorized network nodes, as they can be compromised post-authentication and impede the reliability and resilience of the network. Trust and Reputation Management (TRM) is an effective approach to continuously evaluate the trustworthiness of each participant by collecting and processing evidence of their interactions with other nodes and the infrastructure. In this article, we argue that blockchain-based TRM is critical to build trustworthy 6G networks, where blockchain acts as a decentralized platform for collaboratively managing and processing interaction evidence with the end goal of quantifying trust. We present a case study of resource management in 6G networks, where blockchain-based TRM quantifies and maintains reputation scores by evaluating fulfillment of resource owner's obligations and facilitating resource consumers to provide feedback. We also discuss inherent challenges and future directions for the development of blockchain-based TRM for next-generation 6G networks.
}

%--------------------------------------------------------------------------------------
\section{Introduction}
The future of wireless communication technology, commonly known as the sixth generation (6G) communication network, is envisioned to enable various futuristic technologies, such as hologram and multi-sensory augmented reality, reliable telemedicine, and large-scale interconnected autonomous vehicles~\cite{khan2021}. These future technologies are the backbone of the so called Internet of Everything (IoE), which demands extremely high communication throughput with significantly low latency and support for extensive device heterogeneity. To achieve this ambitious target, researchers are exploring the possibilities of operating in TeraHertz (THz) frequency band and utilization of spatial multiplexing to allow base stations to provide simultaneous access to thousands, if not millions, of ultra-dense wireless connections~\cite{li2020}.

The salient characteristics of envisioned 6G networks bring new challenges in providing resilient communications with more complexities than the previous generations~\cite{nguyen2021a}. 6G will give rise to more distributed Device-to-device (D2D) applications where there will be a greater need for participants to directly interact with other participants, i.e., move away from traditional client-server interactions where the assumption is that the server is trusted. In fact, ultra-dense connections would involve large number of interconnected devices with an absence of pre-established trust between participants. Incorporation of reliable authentication schemes can protect the network from unauthorized adversaries. However, it cannot ascertain in situ reliability and trustworthiness of authorized network nodes, which may become faulty or compromised post-authentication. Compromised devices can behave maliciously, eventually causing severe detrimental effects, as most critical infrastructures are anticipated to be connected in the 6G universe~\cite{ahmad2020}. The trustworthiness of network participants, therefore, becomes an important issue, as the absence of trust may discourage participants from cooperating in the network.

Trust and Reputation Management (TRM) is an effective approach to overcome the aforementioned issues, wherein an authority continuously evaluates the trustworthiness of each participant by collecting and processing feedback and ratings from other network participants. TRM calculates the reputation scores based on collected interaction evidence to quantify the trustworthiness of the participants. The state-of-the-art TRM approaches for 5G~\cite{ahmad2020}, cannot provide reliable mechanisms for future 6G applications, as they have poor attack resiliency, suffer from inefficient reputation score calculation, and are unable to scale. In addition, TRM for 6G-enabled applications should rely on the coordination between network entities and users to collect evidence of misbehavior, which are not supported in the conventional TRM.

Blockchain has been projected to support various aspects of future 6G communications, as predicted by the Federal Communications Commission (FCC)~\cite{nguyen2021a}. Recent work has proposed the notion of blockchain radio access network (B-RAN) for efficient resource management and resource sharing~\cite{wang2021}. Furthermore, blockchain has the potential in building more secure and trustworthy TRM for 6G networks. The decentralized nature of blockchain and its salient features, including transparency, tamper-resilient, verifiability, and smart contract execution, show promise in resolving the aforementioned issues. In this article, we argue that blockchain-based TRM can  create resilient and trustworthy 6G networks. We demonstrate the potential of blockchain through a case study of a resource management scheme for 6G networks where we propose a smart contract-based TRM to provide resilient and efficient resource sharing for network participants. We show how our solution can meet 6G requirements in Figure~\ref{fig:design-consideration}. We also discuss the potential challenges and future directions of blockchain-based TRM for 6G. In summary, this article makes the following contributions:
\begin{itemize}
    \item We motivate the use of blockchain-based TRM for enabling secure and trustworthy 6G networks and discuss the challenges of meeting 6G performance requirements for TRM.
    \item We propose a blockchain-based TRM that aligns with the enabling technologies and performance requirements of 6G. Specifically, we design our TRM to be scalable and resilient against attacks in trust-based systems, and providing efficient trust calculation.
    \item We present a case study of 6G resource sharing to demonstrate the practicality of blockchain-based TRM. We show that the TRM can provide higher resource utilization rate and give distinctive scoring for reliable and unreliable nodes.
\end{itemize}

\section{Background and Motivation}
Trust is a subjective property that exhibits an expectation that a participant would perform actions as predicted, which is gradually built from repeated interactions~\cite{putra2021a}. Similarly, reputation implies an aggregated trust degree from multiple participants that measures how a participant behaves over a temporal horizon. In the context of 6G networks, trusting a network participant means that the participant would perform all protocol/system operations, and a high reputation value indicates that the performance of the participant has been satisfactory according to several other users over a period of time. However, future 6G networks are expected to have a large number of interconnected devices owned by various unknown participants, which may be untrustworthy. There is no guarantee that these devices would correctly follow the pre-defined protocols without acting maliciously.

\begin{figure}
\centering
\includegraphics[width=0.5\textwidth]{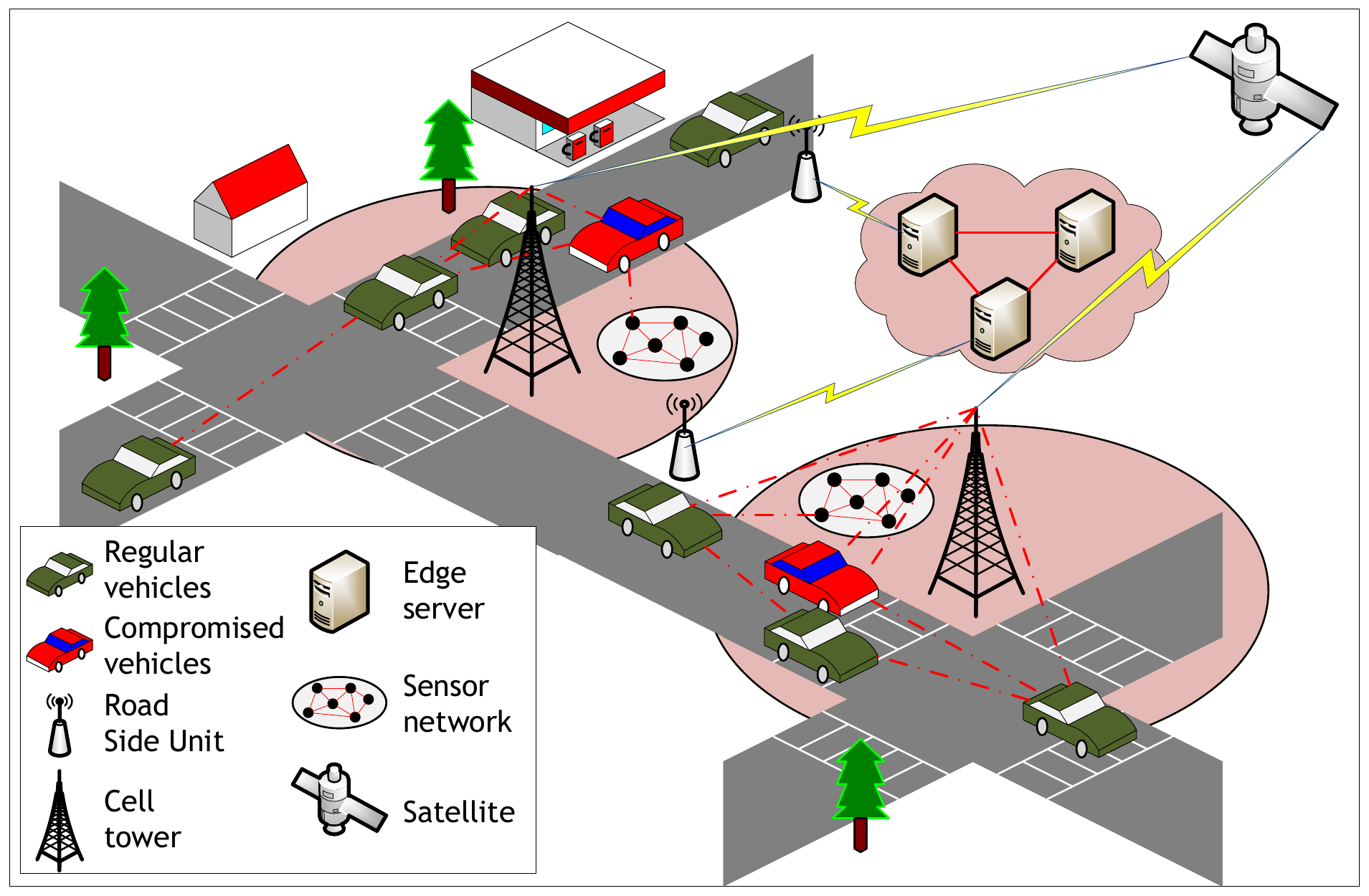}
\caption{An example of TRM architecture for 6G-supported VANET.}
\label{fig:vanet-architecture-diagram}
\end{figure}

TRM is an approach that can address trust issues in 6G networks by quantitatively assessing the trustworthiness of network participants from direct experience or recommendations from other participants. In general, the trust and reputation score can be used as a safeguard to manage the associated risk in interacting with other participants in 6G networks, enabling a trustworthy communication system. We consider an example of 6G-enabled Vehicular Networks (VANET), where there is a greater need to trust other peers or vehicles (Figure~\ref{fig:vanet-architecture-diagram}). Here, TRM would enhance traditional security in VANET, e.g., Public Key Infrastructure (PKI), which is not adequate to provide assurance of the credibility of exchanged messages, as vehicles may be malicious or compromised, e.g., due to faulty on-board sensors. In TRM for 6G VANET, dedicated Road Side Units (RSU) could collect information from neighboring vehicles to validate the exchanged message and assign a score to each vehicle and corresponding messages, using which the TRM can scan for malicious or faulty vehicles. In addition, each vehicle can query the RSU to obtain the latest reputation scores of any vehicle in the proximity.

\begin{table*}
\centering
\caption{Comparison of existing work in TRM for mobile networks.}
\label{tab:comparison}
\scalebox{0.9}{
\begin{tabular}{lccccc}
\toprule
\textbf{Ref.} & \textbf{Architecture} & \textbf{Transparency} & \textbf{Scaling} &  \textbf{Resiliency} & \textbf{Interoperable} \\ 
{} & {} & \textbf{\& Verifiability} & {} & {} & \textbf{TRM} \\ \midrule
\cite{pang2021} & Conventional (CT) & \xmark & Unable to scale & Moderate & \xmark \\
\cite{chouikhi2020} & Conventional (DT) & \xmark & Unable to scale & Low & \xmark \\
\cite{kang2019a} & Blockchain (DC) & \checkmark & Moderate & Moderate & \xmark \\
\cite{ye2022} & Blockchain (DC) & \checkmark & Moderate & Good & \xmark \\
This article & Blockchain (DC) & \checkmark & Good & Good & \checkmark \\
\bottomrule \noalign{\smallskip}
\multicolumn{6}{c}{\small{CT = Centralized, DT = Distributed, DC = Decentralized.}}
\end{tabular}
}
\end{table*}

\subsection{State of the art}
In this section, we compare our solution against existing conventional and blockchain-based TRM approaches, summarized in Table~\ref{tab:comparison}.

\subsubsection{Conventional TRM}
Conventional TRM can be broadly classified into centralized and distributed TRM. In centralized TRM, a Trusted Third Party (TTP) is responsible to assess the trustworthiness of each network participant, after which the TTP calculates and assigns trust scores according to collected collaboration evidence, such as ratings and feedback. For instance, in a centralized TRM for Smart Driving Vehicles (SDV)~\cite{pang2021}, edge servers controlled by a TTP coordinate the reputation management by collecting vehicle information and generating vehicle reputation table. The vehicle reputation scores are derived from the past history of completing tasks accurately and on time. The scores are used in selecting the relay vehicles for Vehicle-to-Everything (V2X) communications. However, relying on a TTP for managing a TRM poses serious risks. For instance, when the TTP is faulty or compromised, there is no guarantee that the underlying trust calculation would remain intact, as there is no assurance of the integrity and safety of the data. In addition, other network nodes cannot verify the trust calculation carried out by the TTP.

On the other hand, distributed TRM relies on the collaboration of network nodes, as in a distributed TRM for VANET~\cite{chouikhi2020}. To improve the network efficiency and protect the network from adversaries, each vehicle in the network is assigned a score computed by its neighboring vehicles as per their interaction experience. These scores are utilized to assess the veracity of incoming application messages that may be assessing the road condition, e.g., traffic and road events, and identify misbehaving nodes. However, malicious vehicles may attempt to gain advantage by spoiling honest vehicles' reputation or colluding to improve their scores, as this scheme has poor transparency. In addition, it is difficult to scale the approach, especially when the number of nodes increases exponentially.

\begin{figure*}
    \centering
    \includegraphics[width=\textwidth]{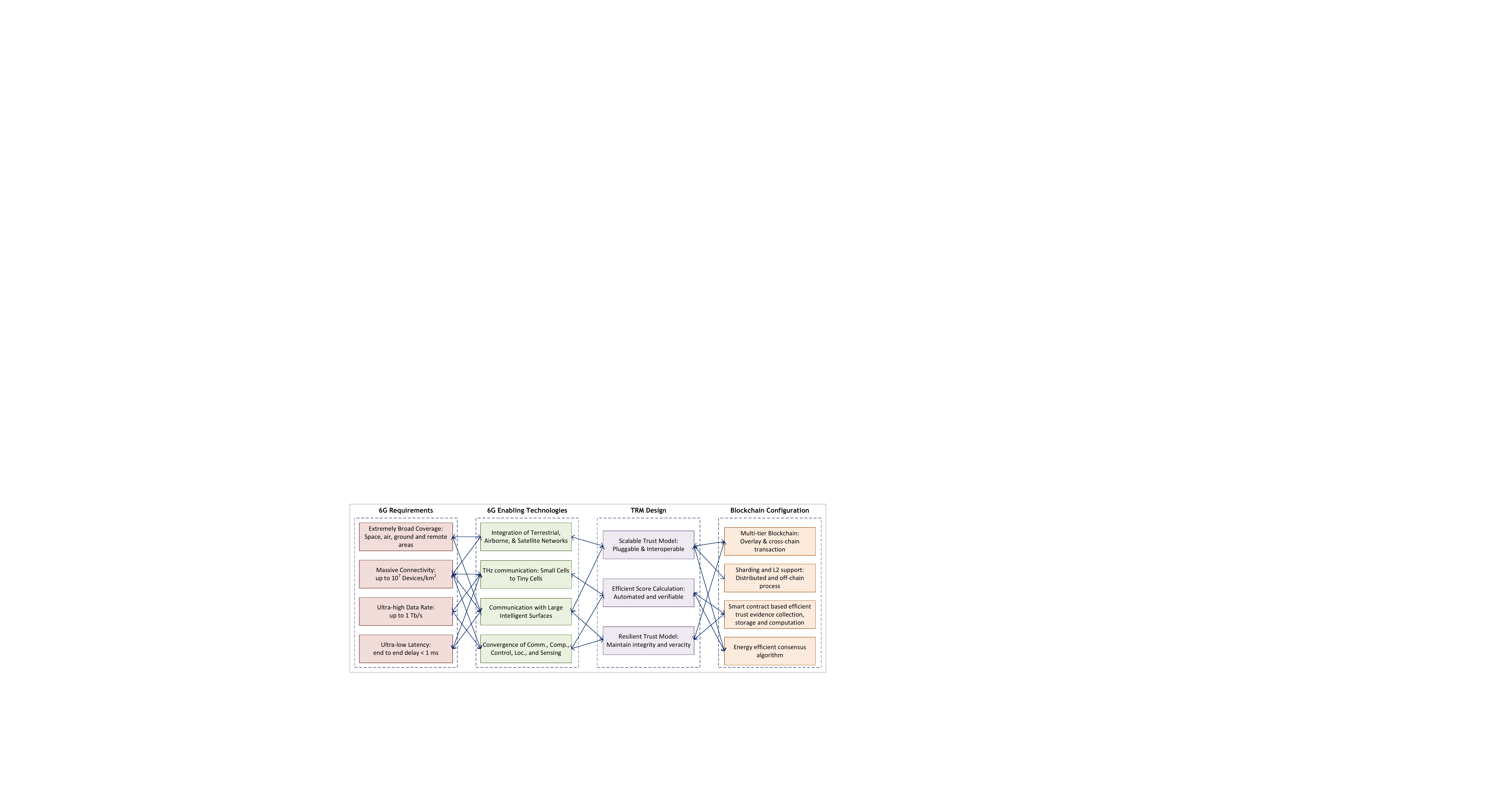}
    \caption{6G Requirements and the improvements of TRMS with the help of blockchain.}
    \label{fig:design-consideration}
\end{figure*}

\subsubsection{Motivation for Blockchain}
The risks and challenges in conventional TRM motivate the use of blockchain for TRM. Blockchain shows promise in resolving the aforementioned issues due to the following salient features.

\color{black}
\textbf{Decentralization.}
The decentralization inherent in blockchain eliminates the reliance on a TTP. The underlying consensus algorithms, e.g., Proof-of-Work (PoW) and Proof-of-Stake (PoS), protect the blockchain from adversaries and malicious insertions.

\textbf{Smart Contracts.}
Blockchain utilizes smart contracts to automate a trusted and deterministic business logic execution~\cite{wang2019b}. In blockchain-based TRM, smart contracts are employed to enforce trustworthy collection of collaboration evidence and trust calculation.

\textbf{Distributed Immutable Ledger.}
In blockchain-based TRM, trust related data is stored on the shared immutable ledger, maintaining high availability. Here, each blockchain node with sufficient resources, e.g., Mobile Network Operator (MNO), can maintain a copy of the ledger, while nodes with limited resources, e.g., UE, may opt to store the block headers and request other data on demand.

\textbf{Pseudonymity.}
Blockchain employs elliptic-curve cryptography, which uses pseudonyms rather than real-world identities for identification purposes. In blockchain-based TRM, each participant is identifiable by its public key, enhancing privacy preservation.

\textbf{Cryptocurrency.}
Conventional mechanisms employ inefficient external processes to manage financial transactions. On the other hand, blockchain readily supports cryptocurrency, which can provide an efficient mechanism for actioning incentives and penalties and to conveniently provide payments.

There have been numerous studies on blockchain-based TRM in different applications, such as supply chain management, crowdsourcing, and VANET. These studies investigate how the above mentioned features of blockchain contribute to the development of a decentralized TRM, with the aim of providing end-to-end trust between nodes in the network. For instance, in~\cite{kang2019a} the authors show that VANET can benefit from blockchain-based TRM, where reputation is used in PoS consensus to select reliable miners, i.e., RSU. Here, each vehicle updates the reputation of miners based on its past experience in interacting with the miners and recommendations from other vehicles. Only highly reputed miners are selected to join the PoS consensus process. The TRM can protect the network from voting collusion resulting in secure blockchain-enabled VANET. However, this scheme~\cite{kang2019a} does not maximize the use of blockchain, as smart contracts are not utilized. In~\cite{ye2022}, a smart contract-based TRM is proposed to secure and improve spectrum sensing in dynamic spectrum access for 5G cognitive radio. Here, smart contracts realize the automation of trust evidence collection, i.e., sensing results, and computation of trust scores, which are determined from sensing nodes' performance in cooperative sensing.

However, these approaches cannot fulfill the novel challenges in providing secure blockchain-based TRM for 6G networks. First, given the wide coverage of 6G which includes both terrestrial and non-terrestrial networks, the existing TRM approaches cannot be directly applied as they are unable to scale. Second, 6G demands massive connectivity which foresees more than millions of devices in a particular area. As such, the TRM should incorporate efficient trust model and calculation, which are overlooked in the existing approaches. Lastly, the large number of interconnected untrustworthy devices may make existing protection approaches against TRM attacks impractical. We aim to address these challenges in the following section.

%--------------------------------------------------------------------------------------
\section{Our Proposed Solution}
In this section, we present a case study and evaluations to demonstrate the benefits of incorporating blockchain-based TRM in 6G communications. We consider a scenario of dynamic resource sharing, which is expected to be the one of the core features of 6G technology. As explained earlier, we argue that proper re-design of TRM and configuration of blockchain components are required to fulfill 6G requirements as seen in Figure~\ref{fig:design-consideration}.

\subsection{Dynamic Resource Sharing in 6G}
6G applications demand pervasive intelligent services to bring intelligence closer to the end devices, which requires extensive utilization of scattered computing resources~\cite{hu2021}. However, there is a finite amount of available computing resources in the network, which requires an efficient mechanism to achieve maximum utilization of scarce resources. In 6G networks, resource management plays an important role in managing the resource pool by means of dynamic resource allocation, e.g., allocating idle computing resources. Note that both nodes that share a resource should agree on pre-determined obligations and a Service Level Agreement (SLA). Here, blockchain can act as a trusted intermediary to allocate the resources dynamically through a secure, auditable, and transparent process without relying on a TTP. For instance, smart contracts can manage resource allocation in a verifiable manner, while the ledger can transparently store allocation related information, e.g., resource allocation, resource lease mappings, and pre-determined SLA or obligations.

While blockchain can provide assurance of secure resource allocation, it alone cannot ascertain that the participants adhere to apportioned resources. There might be several participants that do not fulfill their obligations and do not conform to the pre-determined SLA, making the dynamic resource allocation inefficient. We propose a blockchain-based TRM scheme to build secure and trustworthy resource management in 6G networks. We show how TRM can facilitate secure resource management in 6G where trust and reputation scores play an important role in characterizing reliable, faulty and unreliable resource owners.\\

\begin{figure*}
\centering
\includegraphics[width=\textwidth]{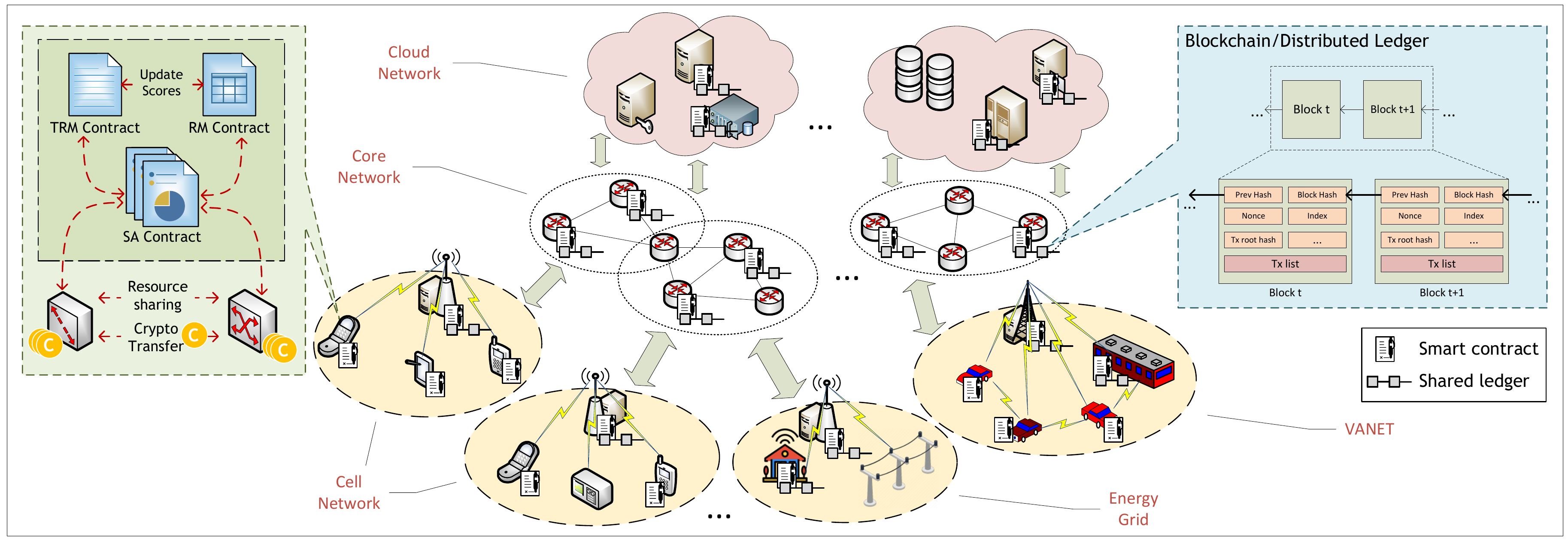}
\caption{Illustration of the proposed blockchain-based TRM for resource sharing in 6G networks.}
\label{fig:architecture-diagram}
\end{figure*}

\subsection{TRM for 6G Networks}
Our proposed blockchain-based TRM architecture consists of several distributed entities, such as edge servers, small cell Base Stations (BS), UE, and IoT devices (Figure~\ref{fig:architecture-diagram}). In general these nodes are divided into Resource Owners (RO), e.g.,, edge servers and MNO, and Resource Users (RU), e.g., UE and IoT devices. These nodes collaboratively form a consortium blockchain, where all participants should be known in advance but not necessarily trusted. In this consortium blockchain, each participant is identifiable by a public-key linked to a cryptocurrency account.

Multiple vision papers note that the 6G requirements would be driven by novel enabling technologies and trends, such as Large Intelligent Surfaces (LIS), incorporation of Non-Terrestrial Networks (NTN), and Convergence of Communications, Computing, Control, Localization, and Sensing (3CLS)~\cite{saad2020,zhang2019}. We argue that TRM should also be re-designed to fit with these new requirements and technologies, wherein blockchain would play a significant supporting role. We anticipate the need for proper configuration of the blockchain architecture by selecting appropriate underlying components, such as the consensus algorithm. We show in Figure~\ref{fig:design-consideration} how our solution can fulfill these novel requirements and addresses the weaknesses of the previous work. \\

\noindent \textbf{Scalable Trust Model}\\
6G networks will integrate satellite, airborne and terrestrial networks to provide wider coverage area and massive connectivity. In addition, 6G would implement LIS to provide support to traditional massive MIMO~\cite{saad2020}, resulting in a network with bigger scale. 6G-enabled TRM, consequently, needs to be designed with scalability in mind. First, TRM should be designed in a hierarchical way accommodating several autonomous sub-networks. Note that, these networks may have their own independent trust models. Second, these autonomous TRM should be interoperable, as nodes may move from one sub-network to another. TRM can employ a mechanism allowing the transfer of trust scores between sub-networks. For instance, a pluggable trust model would allow each sub-network to implement its own trust model enhancing interoperability.

Here blockchain would enable connections between sub-networks by employing a hierarchical blockchain structure, in which sub-networks may have their own autonomous blockchain. To provide interoperability, cross-chain transactions may be incorporated which would guarantee immutability and auditability~\cite{robinson2022}. 
In addition, blockchain could employ sharding mechanism, where the blockchain is partitioned to distribute the workloads to different nodes. As such, each node only maintains information relevant to their shard. Off-chain scaling, referred to as Layer 2 (L2) solution, is another alternative, which works by building an overlay network or application that runs on top of the main blockchain (Layer 1), inheriting security and integrity guarantees of the underlying blockchain. \\

\noindent
\textbf{Efficient Score Calculation}\\
Tiny cells and communication with LIS in 6G would result in more connected devices with more complexity, bringing more challenges to TRM especially in calculating a large number of trust and reputation scores. 6G-enabled TRM should thus incorporate efficient trust and reputation score calculation in an automated and verifiable manner. Most conventional trust models re-calculate the scores by iterating through the full history of trust evidence and require all nodes in the network to perform these calculations, which results in unnecessary overheads. In 6G-enabled TRM, we optimize the trust model with a simple recursive computation to re-calculate the scores. We also fully offload the collection of trust evidence and calculation of trust scores to smart contracts, which provide automation and verifiability through a single blockchain transaction. In addition, implementing an efficient score calculation requires careful selection of consensus algorithms. We thus recommend incorporating resource efficient PoS instead of resource consumptive PoW.

We utilize a hierarchical structure of three smart contracts: (i) TRM, (ii) Resource Manager (RM), and (iii) Sharing Agreement (SA) contracts. First, \textbf{TRM contract} is one of the two parent contracts responsible for keeping track of each participant's trust and reputation score, together with the trust related data such as evidence of malicious behaviors or feedback from other participants. TRM contract also maintains the updated trust and reputation scores. Second, \textbf{RM contract} is another parent contract that hosts the resource bidding data, which includes the detail of listing, allocation and release of the available resources. Lastly, \textbf{SA contract} is a child contract, which is created every time two participants agree to share computing resources. SA contract enforces the pre-determined SLA during resource bidding, which covers the terms and obligations. SA contract records any undesirable incidents and reports them to TRM and RM contracts. SA contract is temporary and is marked as obsolete upon completion of the resource sharing. \\

\begin{figure*}
    \centering
    \begin{subfigure}[b]{0.43\textwidth}
        \includegraphics[width=\textwidth]{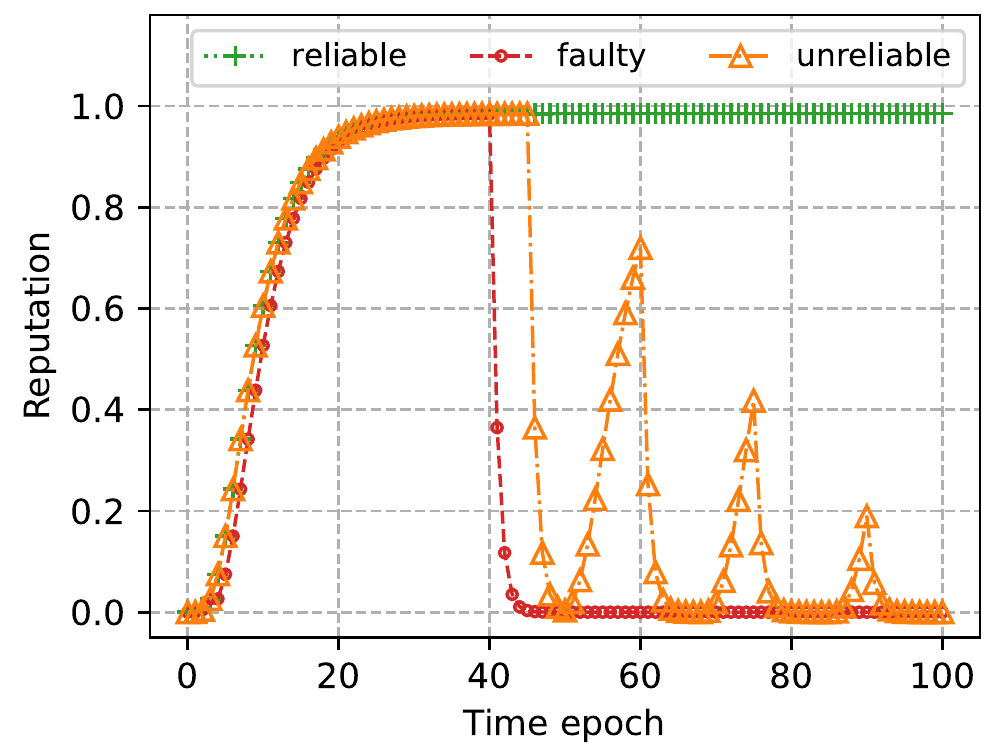}
        \caption{The evolution of reputation scores for a reliable, faulty, and unreliable resource owner.}
        \label{fig:reputation-evolution}
    \end{subfigure}
    ~
    \begin{subfigure}[b]{0.43\textwidth}
        \includegraphics[width=\textwidth]{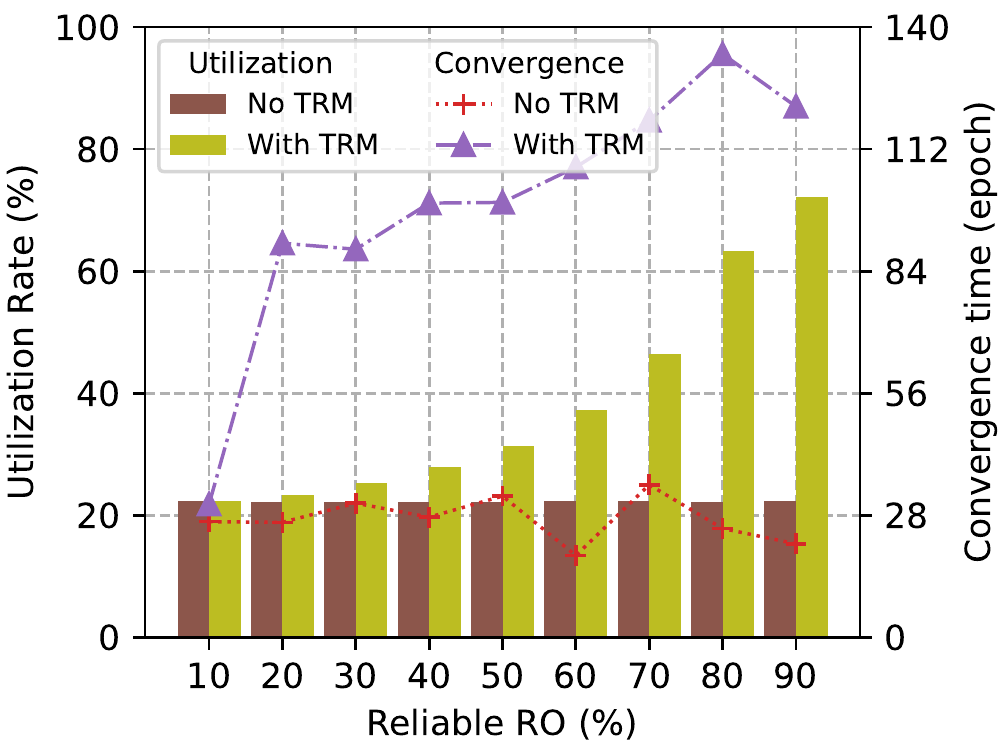}
        \caption{Utilization rate of the resource under different proportions of reliable RO.}
        \label{fig:resource-utilization}
    \end{subfigure}
    \caption{Simulation results in resource sharing for edge computing in 6G network.}
    \label{fig:simulation-results}
\end{figure*}

\noindent \textbf{Resilient Trust Model}\\
Communications with LIS and convergence of 3CLS would bring new challenges in providing secure connectivity. 6G-enabled TRM should also be designed with resiliency in mind to ensure robustness against attack in trust-based systems. With the help of blockchain's salient features, such as decentralization and smart contracts, TRM can be made resilient against these attacks, as explained below.
\begin{itemize}
    \item \textbf{Self-promoting attacks:} In this type of attack, adversaries try to illegitimately increase their trust and reputation scores by providing positive feedback to themselves. To overcome this attack, the trust model should take into account the source of the rater when calculating the scores. As such, consecutive feedback would only be regarded as a single feedback and would not increase the trust scores. In addition, smart contracts can also detect repetitive transactions coming from the same address and reject the additional transactions to prevent this attack.
    \item \textbf{Bad-mouthing attacks:} As opposed to self-promoting, adversaries attempt to denounce honest nodes' trust scores by submitting erroneous negative feedback. To defend against this attack, the TRM could require raters to provide trust evidence when submitting feedback.
    \item \textbf{Ballot-stuffing attacks:} In this attack, an adversary colludes with other nodes to illicitly boost its reputation score. With the blockchain integration, the TRM can require the raters to provide escrow fee when submitting feedback, discouraging the adversaries to launch this attack. In addition, escrow fee can also be used during node registration, which would help discourage Sybil attacks, where an adversary creates multiple fake identities to their benefit, e.g., illegitimately increasing their reputation scores.
    \item \textbf{Whitewashing or newcomer attacks:} To remove its malicious track records, an adversary launches this attack by rejoining the network with a fresh identity. However, the TRM may detect this attack during registration, as each node has to provide real attributes which cannot be easily changed, e.g., device ID, hence the registration would fail.
    \item \textbf{On-off attacks:} In this attack, adversaries alternate between providing good and bad service to different target nodes, hoping to escape detection if the bad service is not picked up as being malicious by some nodes. However, our reputation scores present a global view of each node, meaning bad behavior would be noticed by other nodes as they can view the public record on-chain. In addition, the trust model enforces rigorous punishments against bad behaviors, as a single bad experience would result in abrupt decline of the reputation score.
\end{itemize}

\begin{table}
\centering
\caption{Simulation details.}
\label{tab:simulation-details}
\scalebox{1}{
\begin{tabular}{ll}
\toprule
\textbf{Parameter} & \textbf{Value} \\ \midrule
Number of RO & 500 \\
Number of RU & 600 \\
Node type & [reliable, unreliable, faulty] \\
Reliable RO & [10\%, 20\%, $\ldots$, 90\%]\\
Trusted RO & 10\% \\
Lenient RU & 15\% \\
Iteration & 100 \\
\bottomrule
\end{tabular}}
\end{table}

\subsection{Experimental Evaluation}
We performed two experiments, to measure the evolution of the reputation score and utilization rate for demonstrating the practical role of blockchain-based TRM in 6G resource sharing. We consider an edge computing scenario in a small cell 6G sub-network, where resource constrained UE and IoT devices, i.e., RU, offload computing tasks to more powerful nodes, such as edge servers and MNO, i.e., RO. Here, the TRM continuously evaluates the trustworthiness of each participant and calculates reputation scores accordingly. We consider a blockchain-based setup similar to~\cite{putra2021a} and summarize important simulation parameters in Table~\ref{tab:simulation-details}.

In the first evaluation, we consider a scenario of three RO with different characteristics, namely (i) reliable, (ii) faulty, i.e., stops functioning at one point, and (iii) unreliable, i.e., provides unstable service. In our experiment, we configure all ROs to provide reliable services up to 40 interactions. The faulty and untrustworthy RO are configured to become unreliable and unstable, respectively, after 40 interactions. We plot the reputation score evolution in Figure~\ref{fig:reputation-evolution}. The results demonstrate that our blockchain-based TRM could assign distinctive reputation scores to each node according to how they attain the pre-determined SLA. The reliable RO can maintain high reputation scores, while both the faulty and unreliable RO experience significant reputation score drops once they show inconsistent behavior. In practice, these scores can help RU pick reliable RO by looking at the interaction history manifested in the reputation scores.

In the second evaluation, we investigate the utilization rate of available resources, where we vary the proportion of reliable RO in the network under two conditions, i.e., with and without a TRM in place. We plot in Figure~\ref{fig:resource-utilization} the utilization rate and the convergence time (epoch) at which the network achieves maximum resource utilization. The results indicate that the utilization rate of the sub-network without TRM remains low, approximately around 20\%, with stable convergence rate of around 28 epoch, regardless of the proportion of reliable RO. The results confirm that RU are only inclined in offloading computing tasks to well-known publicly trusted RO and are more reluctant to offload computing tasks to RO which they do not trust. We refer to a RU which is willing to offload tasks to untrusted RO as lenient RU, which constitutes only up to 15\% of the total RU (see Table~\ref{tab:simulation-details}). While the convergence time increases for the sub-network with TRM, the network achieves higher utilization rate as the proportion of reliable RO grows. Here, we argue that TRM would help RU identify reliable RO in the network. As more positive and satisfactory interactions happen, RO gains higher reputation scores, which increases the trust in RO, resulting in increased utilization rate over time.

\subsection{Summary}
The following points summarize our blockchain-based TRM solution and evaluations:
\begin{itemize}
    \item In 6G-enabled edge computing, the amount of available computing resources are limited, which triggers the need of resource sharing to perform computing tasks. However, most nodes are reluctant to join this sharing scheme, as 6G does not guarantee the trustworthiness of the nodes in the network. Here, blockchain-based TRM is an effective approach to encourage resource sharing by assessing the trustworthiness of 6G nodes in a decentralized and verifiable manner.
    \item However, due to the novel requirements inherent to 6G, direct adoption of existing TRM approaches is not feasible, as proper adjustments are required for both the TRM and the underlying blockchain.
    \item The experiment results show that our TRM can apply distinctive scores to reliable and unreliable nodes and increase the utilization of available resources in the network.
\end{itemize}

%--------------------------------------------------------------------------------------
\section{Challenges and Future Directions}
In this section, we briefly discuss the challenges and possible future directions for research in blockchain-based TRM for 6G networks.

\textbf{Security and Privacy.}
Future research should provide rigorous measures for security and privacy preservation entailing the TRM model and blockchain design. Unprecedented scale in 6G network would enlarge the attack surface which may render the conventional defense strategy for TRM attacks infeasible. While blockchain offers pseudonymity, the transparent nature of the shared ledger may allow adversaries to launch de-anonymization attacks, increasing the risk of privacy infringement. As such, TRM in 6G should always be designed with security in mind and also with adequate measures for privacy preservation, which can be addressed by using privacy-preserving TRM models. In addition, as the complex nature of smart contracts has provoked several attacks, special attention to the security of smart contracts is necessary, for instance via formal verification.

\textbf{Scalability.}
Massive scale of 6G network demands further research in scalability of blockchain-based TRM in all aspects, which is a non-trivial task. While advancements in hardware and infrastructure would help increase the network capacity, research in software-based blockchain design solutions, for instance via sharding, sidechains, efficient consensus algorithms, and hierarchical blockchains, should be investigated. In addition, the scalability of the TRM model can also be improved, for instance by allowing an aggregated interaction evidence to be stored instead of redundant trust information that may overload the network in the long run.

\textbf{Interoperability.}
6G will provide wider connectivity by integrating autonomous networks, some of which may have their own independent TRM. This highlights the need for reliable interoperability between networks, which requires further investigation. More research should be undertaken to investigate how a reliable and secure reputation score transfer between independent TRMs can be realized, which requires rigorous assessment of distinctive trust metrics in different autonomous networks. To help realize this, further research should provide seamless interoperability between blockchains, allowing for cross-chain data transfer.
\color{black}

\textbf{Artificial Intelligence.}
As Artificial Intelligence (AI) is envisioned to be an integral part of 6G, further research should aim for using AI-enabled 6G to enhance TRM with the help of decentralized nature of blockchain. For instance, future work may explore novel AI-based reputation models for TRM. AI would enable robust reputation score calculation from large and complex node interaction evidence, which would be more resilient to sophisticated attacks in TRM. Here, blockchain can be utilized to execute federated learning for collecting scattered trust evidence in various autonomous networks in 6G.

\textbf{Post-quantum cryptography.}
There is an urgent need to conduct research on quantum-resistant cryptography to replace current security standards in blockchain systems, as it is expected that by the time 6G has reached commercial readiness, quantum computers will concurrently be available in the market. Quantum computers have the potential to decipher current  encryption standards, which is previously unfeasible. While symmetric cryptography, e.g. hash functions, is not significantly affected by quantum computation, public-key cryptography which is widely used in current blockchain systems is under significant threat. Thus, there is an urgent need to research post-quantum cryptography to guarantee security and reliability of blockchain-based TRM.

%----------------------------------------------------------------------------------------
\section{Conclusion}
In this article, we proposed a blockchain-based TRM to secure and enable efficient utilization of network resources in 6G communications. We first discussed the benefits of incorporating TRM in 6G communications and motivated the use of blockchain-enabled TRM to remove the potential risks of conventional approaches. As a use case, we introduced an architecture of blockchain-based TRM for resource management in 6G networks. We further elaborated on several challenges and future directions for research in blockchain-enabled TRM for 6G networks.

\printbibliography[title={References}]

\section*{Acknowledgment}
This work was supported by the Cyber Security Research Centre Limited through the Australian Government’s Cooperative Research Centres Programme.

\section*{Biographies}
\begin{small}
\begin{description}
    \item[Guntur Dharma Putra] received his bachelor degree in Electrical Engineering from Universitas Gadjah Mada, Indonesia, 2014. He received his master's degree in Computing Science from the University of Groningen, the Netherlands, 2017. He is currently a Ph.D. candidate at the School of Computer Science and Engineering, University of New South Wales (UNSW) Sydney, Australia. His research interest covers blockchain applications for securing IoT. Guntur is a student member of the IEEE. \\
    \item[Volkan Dedeoglu] is currently a research scientist in the Distributed Sensing Systems Group of CSIRO Data61. His current research focuses on data trust and blockchain-based IoT security and privacy. Volkan also holds Adjunct Lecturer positions at UNSW Sydney and QUT. He completed his PhD in Telecommunications Engineering from University of South Australia in 2013. He obtained MSc in Electrical and Computer Engineering from Koc University (2008), BSc in Electrical and Electronics Engineering from Bogazici University (2006), and B.A. in Public Administration from Anadolu University (2008). \\
    \item[Salil S Kanhere] received his M.S. degree and Ph.D. degree from Drexel University in Philadelphia. He is a Professor of Computer Science and Engineering at UNSW Sydney, Australia. He is a Senior Member of the IEEE and ACM, a Humboldt Research Fellow and an ACM Distinguished Speaker. He serves as the Editor in Chief of the Ad Hoc Networks journal and as Associate Editor of IEEE Transactions on Network and Service Management, Computer Communications and Pervasive and Mobile Computing. He has served on the organising committee of many IEEE/ACM international conferences including PerCom, CPS-IOT Week, MobiSys, WoWMoM, MSWiM, and ICBC. \\
    \item[Raja Jurdak] received the MS degree and the PhD degree from University of California, Irvine. He is a Professor of Distributed Systems and Chair in Applied Data Sciences at Queensland University of Technology, and Director of the Trusted Networks Lab. He previously established and led the Distributed Sensing Systems Group at CSIRO’s Data61. He also spent time as visiting academic at MIT and Oxford University in 2011 and 2017. His research interests include blockchain, IoT, trust, mobility and energy-efficiency in networks. Prof. Jurdak has published over 230 peer-reviewed publications, including two authored books most recently on blockchain in cyberphysical systems in 2020. His publications have attracted over 10,800 citations, with an h-index of 47. He serves on the editorial board of Ad Hoc Networks, Nature Scientific Reports, and on the organising and technical program committees of top international conferences, including Percom, ICBC, IPSN, WoWMoM, and ICDCS. He was TPC co-chair of ICBC in 2021. He is a conjoint professor with the University of New South Wales, and a senior member of the IEEE.
\end{description}
\end{small}

\end{document}

%% file: assets/structure.tex
\usepackage[english]{babel} % English language hyphenation

\usepackage{microtype} % Better typography

\usepackage{amsmath,amsfonts,amsthm} % Math packages for equations

\usepackage[svgnames]{xcolor} % Enabling colors by their 'svgnames'

\usepackage[hang, small, labelfont=bf, up, textfont=it]{caption} % Custom captions under/above tables and figures

\usepackage{booktabs} % Horizontal rules in tables

\usepackage{lastpage} % Used to determine the number of pages in the document (for "Page X of Total")

\usepackage{graphicx} % Required for adding images

\usepackage{enumitem} % Required for customising lists
\setlist{noitemsep} % Remove spacing between bullet/numbered list elements

\usepackage{sectsty} % Enables custom section titles
\allsectionsfont{\usefont{OT1}{phv}{b}{n}} % Change the font of all section commands (Helvetica)

\usepackage{subcaption}

\usepackage{pifont}% http://ctan.org/pkg/pifont
\newcommand{\xmark}{\ding{55}}%

\usepackage{geometry} % Required for adjusting page dimensions

\usepackage[T1]{fontenc} % Output font encoding for international characters
\usepackage[utf8]{inputenc} % Required for inputting international characters

\usepackage{XCharter} % Use the XCharter font

\usepackage{fancyhdr} % Needed to define custom headers/footers
\fancypagestyle{firstpage}{ % Page style for the first page with the title
	\fancyhf{}
}

\newcommand{\authorstyle}[1]{{\large\usefont{OT1}{phv}{b}{n}\color{Black}#1}} % Authors style (Helvetica)

\newcommand{\institution}[1]{{\footnotesize\usefont{OT1}{phv}{m}{sl}\color{Black}#1}} % Institutions style (Helvetica)

\usepackage{titling} % Allows custom title configuration

\newcommand{\HorRule}{\color{Gray}\rule{\linewidth}{1pt}} % Defines the gold horizontal rule around the title

\pretitle{
	\vspace{-30pt} % Move the entire title section up
	\HorRule\vspace{10pt} % Horizontal rule before the title
	\fontsize{20}{24}\usefont{OT1}{phv}{b}{n}\selectfont % Helvetica
	\color{Black} % Text colour for the title and author(s)
}

\posttitle{\par\vskip 15pt} % Whitespace under the title

\preauthor{} % Anything that will appear before \author is printed

\postauthor{ % Anything that will appear after \author is printed
	\vspace{10pt} % Space before the rule
	\par\HorRule % Horizontal rule after the title
	\vspace{20pt} % Space after the title section
}

\usepackage{lettrine} % Package to accentuate the first letter of the text (lettrine)
\usepackage{fix-cm}	% Fixes the height of the lettrine

\newcommand{\initial}[1]{ % Defines the command and style for the lettrine
	\lettrine[lines=3,findent=4pt,nindent=0pt]{% Lettrine takes up 3 lines, the text to the right of it is indented 4pt and further indenting of lines 2+ is stopped
		\color{Grey}% Lettrine colour
		{#1}% The letter
	}{}%
}

\usepackage{xstring} % Required for string manipulation

\newcommand{\lettrineabstract}[1]{
	\StrLeft{#1}{1}[\firstletter] % Capture the first letter of the abstract for the lettrine
	\initial{\firstletter}\textbf{\StrGobbleLeft{#1}{1}} % Print the abstract with the first letter as a lettrine and the rest in bold
}

\usepackage[style=ieee,sorting=none]{biblatex}

\usepackage[autostyle=true]{csquotes} % Required to generate language-dependent quotes in the bibliography